\documentclass[utf8]{frontiersFPHY} 

\setcitestyle{square} 
\usepackage{url,hyperref,lineno,microtype,subcaption}
\usepackage[onehalfspacing]{setspace}
\usepackage{graphicx}
\usepackage{amsmath}    
\usepackage{amssymb}    
\usepackage{txfonts}

\newcommand{\rev}[1]{{ #1}}

\def\keyFont{\fontsize{8}{11}\helveticabold }
\def\firstAuthorLast{Lusso \& Risaliti} 
\def\Authors{Lusso Elisabeta\,$^{1,*}$ and Risaliti Guido\,$^{2,3}$}
\def\Address{$^{1}$Centre for Extragalactic Astronomy, Durham University, South Road, Durham, DH1 3LE, UK \\
$^{2}$Dipartimento di Fisica e Astronomia, Universit\`a di Firenze, via G. Sansone 1, 50019 Sesto Fiorentino, Firenze, Italy \\
$^{3}$INAF--Osservatorio Astrofisico di Arcetri, 50125 Florence, Italy
}
\def\corrAuthor{Corresponding Author}

\def\corrEmail{elisabeta.lusso@durham.ac.uk}

\newcommand{\lo}{\log L_{\rm UV}}
\newcommand{\lx}{\log L_{\rm X}}
\newcommand{\fo}{\log F_{\rm UV}}
\newcommand{\fx}{\log F_{\rm X}}
\newcommand{\Lo}{L_{\rm UV}}
\newcommand{\Fo}{F_{\rm UV}}
\newcommand{\Lx}{L_{\rm X}}
\newcommand{\Fx}{F_{\rm X}}

\newcommand{\mbh}{M_{\rm BH}}

\newcommand{\gammax}{\Gamma_{\rm X}}

\newcommand{\Rs}{R_{\rm S}}
\newcommand{\rg}{r_{\rm g}}

\newcommand{\w}{{FWHM}}

\begin{document}
\onecolumn
\firstpage{1}

\title[The physical relation between disc and coronal emission]{The physical relation between disc and coronal emission in quasars} 

\author[\firstAuthorLast ]{\Authors} 
\address{} 
\correspondance{} 

\extraAuth{}

\maketitle

\begin{abstract}
We propose a modified version of the observed non-linear relation between the X-ray (2 keV) and the \rev{ultraviolet} (2500 \AA) emission in quasars (i.e. $L_{\rm X}\propto L_{\rm UV}^{\gamma}$) which involves the full width at half-maximum, $\w$, of the broad emission line, i.e. $L_{\rm X}\propto L_{\rm UV}^{\hat\gamma}~\w^{\hat\beta}$. By analysing a sample of 550 optically selected non-jetted quasars in the redshift range of 0.36--2.23 from the Sloan Digital Sky Survey cross matched with the XMM-Newton catalogue 3XMM-DR6, we found that the additional dependence of the observed $\Lx-\Lo$ correlation on the $\w$ \rev{of the Mg\textsc{II} broad emission line} is statistically significant. Our statistical analysis leads to a much tighter relation with respect to the one neglecting $\w$, and it does not evolve with redshift.
We interpret this new relation within an accretion disc corona scenario where reconnection and magnetic loops above the accretion disc can account for the production of the primary X-ray radiation. For a broad line region size depending on the disc luminosity as $R_{\rm blr}\propto L_{\rm disc}^{0.5}$, we find that $L_{\rm X}\propto L_{\rm UV}^{4/7} ~\w^{4/7}$, which is in very good agreement with the observed correlation. 
\tiny
 \keyFont{ \section{Keywords:} active galactic nuclei, quasar, supermassive black holes, accretion disc, X-ray} 
\end{abstract}

\section{Introduction}
One of the observational evidences for the link between the accretion disc and the X-ray corona in active galactic nuclei (AGN) is given by the observed non-linear correlation between the monochromatic \rev{ultraviolet} luminosity at 2500 \AA\ ($\Lo$) and the one in the X--rays at 2 keV ($\Lx$).
Such relationship (parameterised as $\lx= \gamma\lo + \beta$) exhibits a slope, $\gamma$, around 0.6, implying that optically bright AGN emit relatively less X-rays than optically faint AGN \citep{avnitananbaum79}. The value of the slope in this relation does not depend on the sample selection, the correlation is also very tight ($\sim0.24$ dex, \citealt{2016ApJ...819..154L}), and independent on redshift \citep{vignali03,strateva05,steffen06,just07,green09,2010A&A...512A..34L,2010ApJ...708.1388Y,2012A&A...539A..48M,2012MNRAS.422.3268J}.
Recently, the $\Lx-\Lo$ relationship (or, more precisely, its version with fluxes) has also been employed as a distance indicator to estimate cosmological parameters such as $\Omega_{\rm M}$ and $\Omega_\Lambda$ by building a Hubble diagram in a similar way as for Type Ia supernovae \citep[and references therein]{suzuki12,2014A&A...568A..22B}, but extending it up to $z\sim6$ \citep{2015ApJ...815...33R,2017AN....338..329R}. Yet, the challenge in interpreting such relation on physical grounds is that the origin of the X--ray emission in quasars is still a matter of debate.

Some attempts at explaining the quasar X-ray spectra were based upon the reprocessing of radiation from a non-thermal electron-positron pair cascade (e.g. \citealt{1982ApJ...258..321S, 1984MNRAS.209..175S,1990ApJ...363L...1Z}). Another possibility rests on a two-phase accretion disc model, where a fraction $f$ of gravitational power is dissipated via buoyancy and reconnection of magnetic fields in a uniform, hot ($T_{\rm cor}\sim100$ keV $\sim10^9$ K) plasma close to the cold opaque disc \citep{1991ApJ...380L..51H,1993ApJ...413..507H,1994ApJ...436..599S, 1998MNRAS.299L..15D, 2000A&A...360.1170R}. The scenario for this model is the following. Phase 1 is the optically thick ``cold'' (tens of eV) disc, whilst phase 2 is a hot optically thin plasma located above (and below) the disc. The seed disc photons illuminate the hot tenuous plasma, and a fraction of them is up-scattered to hard X-rays via inverse Compton scattering, providing the main source of cooling of the plasma. About half of these hard X-ray photons is irradiated back to the disc, contributing to its energy balance, whilst the rest escape and are observed. A stable disc-corona system is then in place only if there is a strong coupling between \rev{ultraviolet} and X-ray photons. This model retrieves the photon index slope of the coronal X-ray spectrum (i.e. $\Gamma_{\rm X}\simeq2$), in close agreement with real data. Nonetheless, it also predicts a nearly equal amount of \rev{ultraviolet} and X-ray radiation (i.e. $\gamma=1$), which is not in agreement with the observed correlation. If the corona is not uniform, but a rather patchy medium, and only a fraction ($f$) of the accretion power is released in the hot phase, the resulting $\gamma$ value is $<1$, but one needs to consider a rather arbitrary number of active blobs \citep{1994ApJ...432L..95H}. Magnetic field turbulence has been recognised not only as an additional heating mechanism in the formation of the corona (e.g. \citealt{1979ApJ...229..318G,2002MNRAS.332..165M,2002ApJ...572L.173L}), but also as an efficient means for the transport of the disc angular momentum (e.g. \citealt{2003ARA&A..41..555B}). Yet, the value of the fraction $f$ of gravitational power dissipated in the hot corona (needed for keeping the plasma at high temperatures), and how the coronal physical state depends \rev{on the black hole mass} and disc accretion rate still remain a matter of debate.

In \citet{2017A&A...602A..79L} we outlined a simple but physically motivated, ad-hoc model to interpret the observed correlation between the \rev{ultraviolet} and the X-ray emission in terms of physical parameters such as the black hole mass \rev{($M_{\rm BH}$, here we considered the normalized value $m=M_{\rm BH}/M_\odot$)}, the accretion rate ($\dot{m}=\dot{M}/\dot{M}_{\rm Edd}$, \rev{where $\dot{M}_{\rm Edd}$ is the accretion rate at Eddington}), and the distance to the black hole ($r=R/\Rs$, where $\Rs=2G\mbh/c^2$ is the Schwarzschild radius). Our main aim was to link such relations with observable quantities (i.e. the observed \rev{ultraviolet} and X-ray luminosities), thus obtaining a relation that can be then compared with the data.

\rev{Here, we further analyse the $\Lx-\Lo-\w$ relation with the goal of understanding its physical origin in the context of black hole accretion physics.}

\section{The observed $\Lx-\Lo-\w$ plane}
\label{The plane}
The quasar sample considered by \citet{2017A&A...602A..79L}  is obtained by cross-matching the quasar SDSS catalogue published by \citet{2011ApJS..194...45S} with the serendipitous X-ray source catalogue 3XMM--DR6 \citep{2016A&A...590A...1R}. Filters are also applied in order to select a clean quasar sample where biases and contaminants are minimised, namely: (i) jetted, broad absorption line quasars and sources with high levels of absorption in the optical ($E(B-V)>0.1$) are removed from the sample; \rev{(ii) we selected those quasars with a full-width half-maximum, $\w$, for the Mg\textsc{II} $\lambda$2800~\AA\ higher than 2000 km s$^{-1}$}; (iii) only sources with good X-ray data (i.e. S/N$>$5 in the 0.2--12~keV EPIC band) and low levels of X-ray absorption (i.e. with an X-ray photon index $\gammax>1.6$) are considered; and finally (iii) the {\it Eddington bias} (see also \citealt{2015ApJ...815...33R} and \citealt{2016ApJ...819..154L} for further details) is minimised by including only quasars whose minimum detectable X-ray flux is lower than the expected one in each observation.  
\rev{The FWHM values for the Mg\textsc{II} emission line as well as the ultraviolet 2500\AA\ luminosities are taken from the \citet{2011ApJS..194...45S} catalogue, which have been measured from the continuum/emission line fitting of the SDSS spectra (see their Section~3 for further details).}
The final clean sample is composed of 550 quasars and it is only $\sim$25\% of the initial sample ($\sim$2,100 quasars with both soft 0.5--2 keV and hard 2--12 keV fluxes), resulting from the stringent filters mentioned above. In the future we will refine the treatment of the systematics by considering additional instrumental systematics, and developing large mock simulations to better understand the impact of the Eddington bias on the results. The main aim is to reduce the rejection fraction, thus maximizing the statistics of the final sample.

\begin{figure}[t!]
\begin{center}
\includegraphics[width=8.5cm]{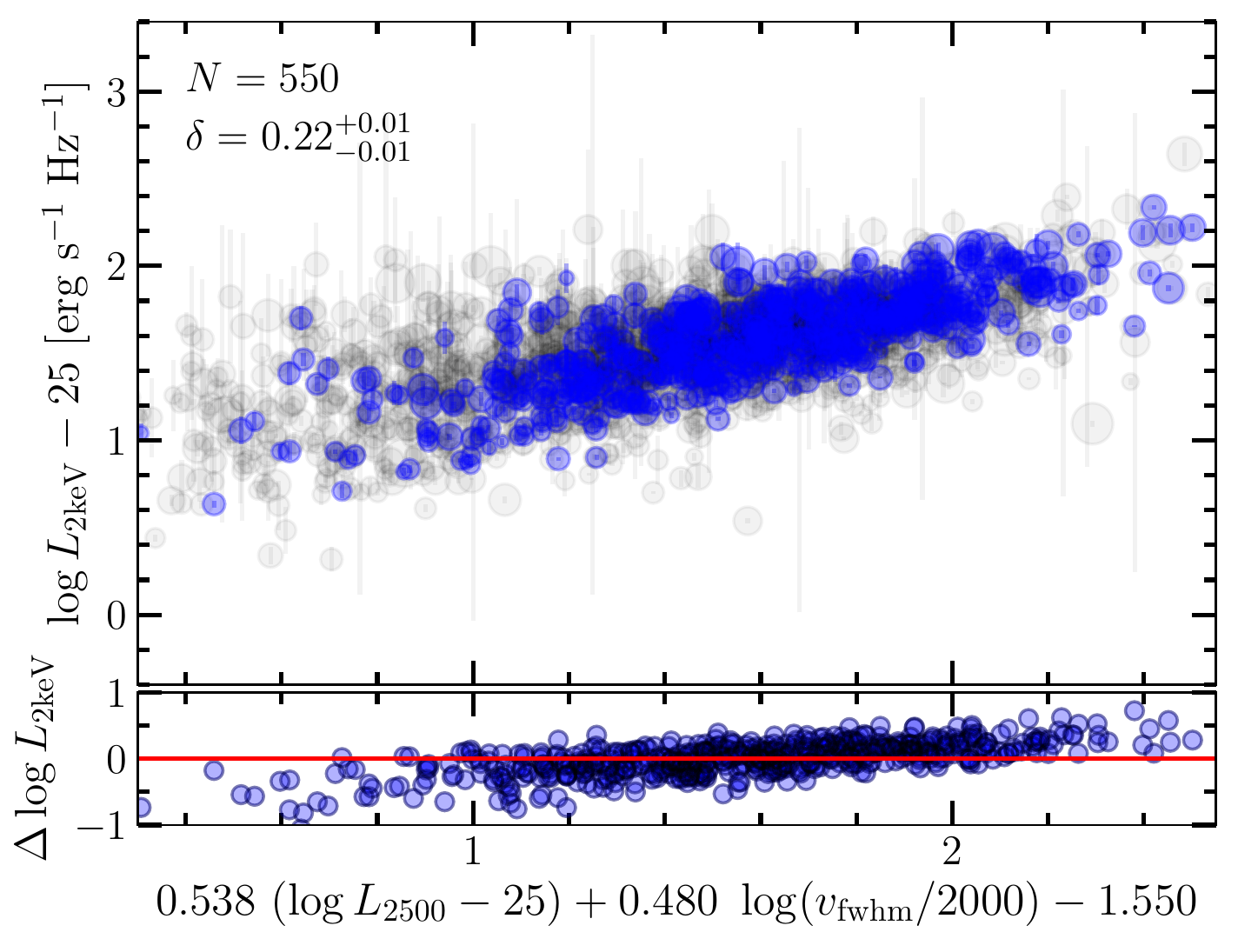}
\includegraphics[width=8.5cm]{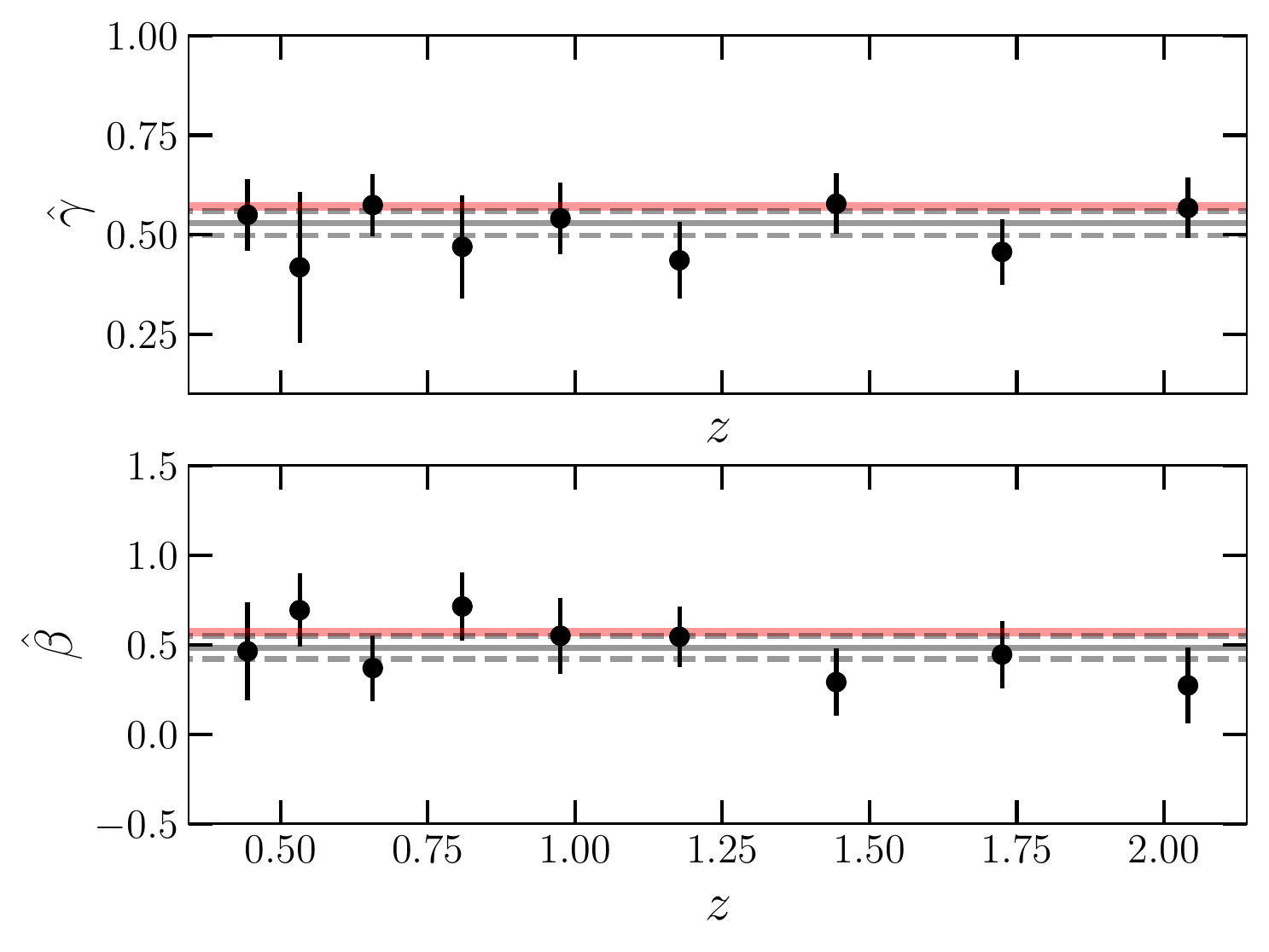}
\end{center}
\caption{ Left: The tight $\Lx-\Lo-\w$ plane (seen edge-on) for the $z<2.3$ SDSS-3XMM-DR6 quasar sample in Lusso \& Risaliti (2017). \rev{The full-width half maximum ($\w$) is computed from the Mg\textsc{II} emission line}. \rev{The size of the points has been rescaled to the $\w$ value, the smaller size corresponds to $\w=2000$ km/s}. The slopes of this correlation are in very good agreement (within the uncertainties) with the ones predicted by our simple (but physically motivated) model, $L_{\rm X}\propto L_{\rm UV}^{\hat\gamma}\w^{\hat\beta}$, where $\hat\gamma=\hat\beta=4/7$ ($4/7\sim$0.571). The quasar sample before applying our filters is also plotted with light grey points. Right: Slopes of the $\Lx-\Lo-\w$ plane as a function of redshift. The red solid line represents the predicted value of $4/7$. The observed $\Lx-\Lo-\w$ plane does not evolve with redshift up to $z\sim2.3$, which is an essential requirement in order to utilize this relation to measure cosmological parameters. The grey solid and dashed lines represent the weighted means and uncertainties of the black points, respectively. Figure based on data derived from Lusso \& Risaliti (2017).}\label{fig:rel}
\end{figure}

Figure~\ref{fig:rel} shows an edge-on view of the $\Lx-\Lo-\w$ plane. The best-fit regression relation for this quasar sample is
\begin{align}
\label{lxlowemcee}
(\log\Lx - 25) = (0.538\pm0.022)(\log \Lo -25) + (0.480\pm0.078)[\log \w - (3+\log2)] + \nonumber\\(-1.550\pm0.122),
\end{align}
with an observed dispersion of 0.22 dex, which is a very tight relation. 
\rev{To fit the data, we adopted {\it emcee} \citep{2013PASP..125..306F}, which is a pure-Python implementation of Goodman \& Weare's affine invariant Markov chain Monte Carlo ensemble sampler.}

We then divided the quasar sample into equally spaced narrow redshift intervals in $\log z$ with a $\Delta\log z <0.1$ to minimise the scatter due to the different luminosity distances within each interval.
We split the sample into 10 intervals with $\Delta\log z =0.08$ and, for each redshift interval, we performed a fit of the $\Fx-\Fo-\w$ relation, $\fx = \hat\gamma_z\fo+\hat\beta_z\w+\hat{K}$, with free $\hat\gamma_z$ and $\hat\beta_z$. The results for the best-fit slopes $\hat\gamma_z$ and $\hat\beta_z$ as a function redshift are shown in Figure~\ref{fig:rel}. Despite the large scatter, both $\hat\gamma_z$ and $\hat\beta_z$ slopes do not show any  significant evolution with time and they are consistent with a constant value in the redshift interval 0.3--2.

\section{The toy model}
\label{The model}

Previous works in the literature suggest that the amount of gravitational energy from the accretion disc released in the hot plasma surrounding the disc itself, likely depends \rev{on the black hole mass $M_{\rm BH}$, the mass accretion rate $\dot{M}$,} and the spin of the black hole. 
In \citet{2017A&A...602A..79L}, the $\Lx-\Lo$ correlation was reproduced if the energy transfer from an optically thick, geometrically thin accretion disc to the corona is confined to the gas-pressure dominated region of the disc. In such simple model, the monochromatic \rev{ultraviolet} and X--ray luminosities show an extra dependence on the $\w$, thus on the black hole mass and accretion rate as $L_{\rm UV}\propto M_{\rm BH}^{4/3} (\dot{M}/\dot{M}_{\rm Edd})^{2/3}$ and $L_{\rm X}\propto M_{\rm BH}^{19/21} (\dot{M}/\dot{M}_{\rm Edd})^{5/21}$, respectively. Assuming a broad line region size function of the disc luminosity as $R_{\rm blr}\propto L_{\rm disc}^{0.5}$ we have that 
\begin{align}
\label{lxlow}
\Lx \simeq 6\times10^4~\Lo^{4/7}~\w^{4/7}~\alpha^{-2/21}  \kappa^{2/7} \times (1-f)^{-6/7}J(r)^{-16/21} ~ {\rm erg~s^{-1} Hz^{-1}},
\end{align}
where $\alpha$ is the standard disc viscosity parameter, $\kappa$ is a calibration constant in the broad line region radius-luminosity relation (i.e. $R_{\rm blr}= k L_{\rm bol}^{0.5}$ \citealt{2015JKAS...48..203T}), and $J(r)=( 1 - \sqrt{3/r})$. 

\begin{figure}[t!]
\begin{center}
\includegraphics[width=13cm]{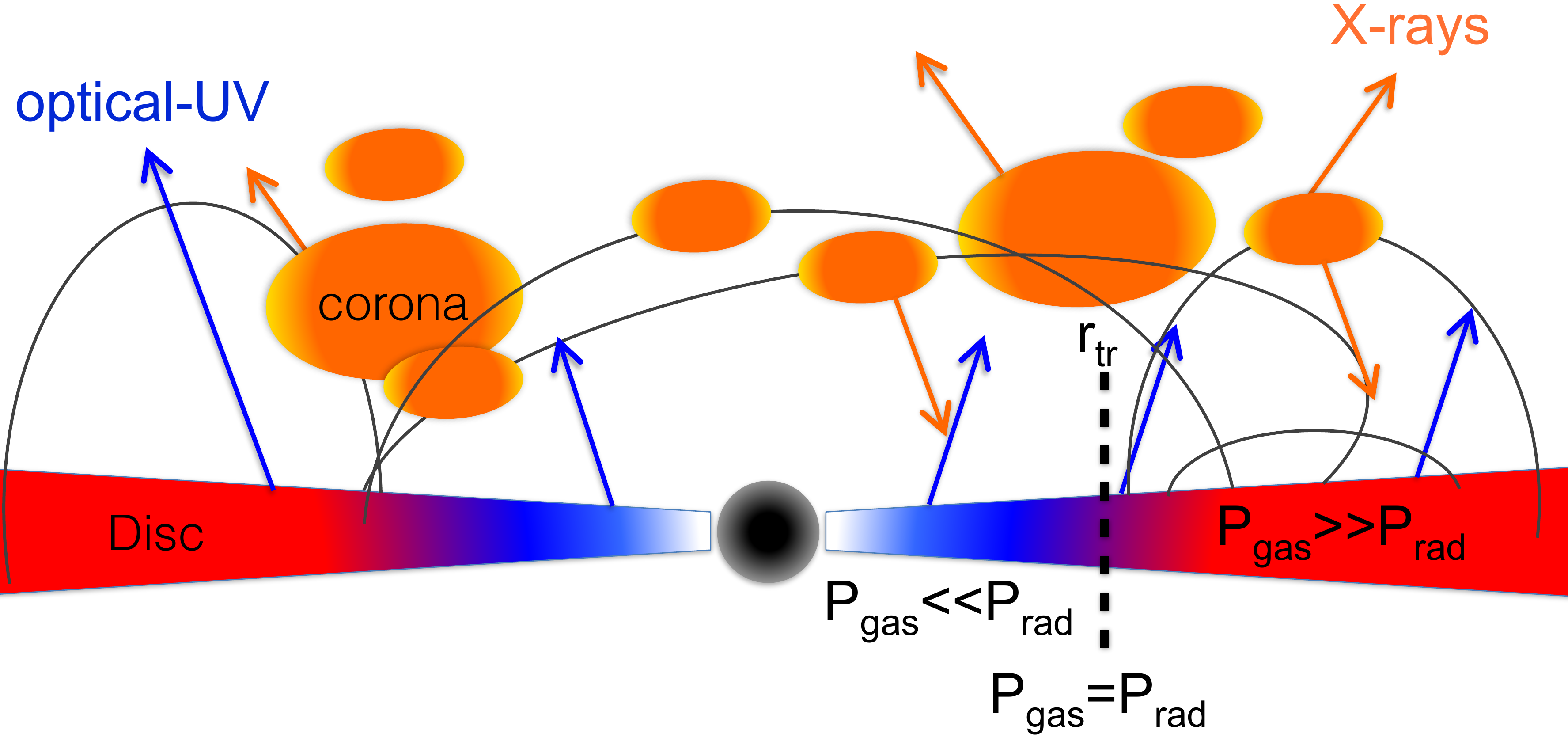}
\end{center}
\caption{A schematic representation of the toy model discussed in \S\ref{The model}. The optically thick, geometrically thin accretion disc emits the seeds photons that illuminate the hot tenuous (possibly clumpy) plasma. A fraction of them is up-scattered to hard X-rays via inverse Compton scattering. Part of these hard X-ray photons is irradiated back to the disc, contributing to its energy balance, whilst the rest escape and are observed. Stable magnetic loop can be formed only in the radiation pressure dominated part of the disc.}\label{fig:mod}
\end{figure}

In this model the bulk of the radiation budget of the X-ray corona is measured at the transition radius ($r_{\rm tr}$) where the gas pressure in the accretion disc equates the radiation pressure, and it is defined as
\begin{equation}
\label{rtr}
r_{\rm tr} \simeq 120 \left( \alpha m \right)^{2/21} \dot{m}^{16/21} (1-f)^{6/7} J(r)^{16/21},
\end{equation}
where $m = \frac{\mbh}{M_\odot}$, and $\dot{m}=\frac{\dot{M}}{\dot{M}_{\rm Edd}}$. The transition radius can vary from a few gravitational radii ($\rg$) to several hundreds $\rg$ for a black hole mass $m=10^{8-9}$, with an accretion rate in the range $\dot{m}=0.1-0.3$, and $\alpha=0.4$. This parameter also depends on the value of $f$ (i.e. the higher is $f$, the smaller is $r_{\rm tr}$). 
Yet, the effective location of the corona may also be placed at small radii of less than tens of $\rg$.
From a qualitative perspective, the torque (caused by the disc rotation) of magnetic field lines (originated at $r\geq r_{\rm tr}$) that connects two opposite sides of the accretion disc (e.g. {\it magnetic reconnection}, a schematic representation of our model is provided in Figure~\ref{fig:mod}) could cause particle acceleration at the magnetic reconnection site, located closer to the black hole (at $r < r_{\rm tr}$), where particles lose their energy radiatively via the inverse Compton process.

\section{Constraining the fraction of accretion power released in the corona}
Our simple (but physically motivated) model also provides constraints on the fraction of accretion power released in the hot plasma in the vicinity of the accretion disc, $f$, as a function the broad line region size, from the observed normalization of $\Lx-\Lo-\w$ plane.

\begin{figure}[t!]
\centering\includegraphics[width=10cm]{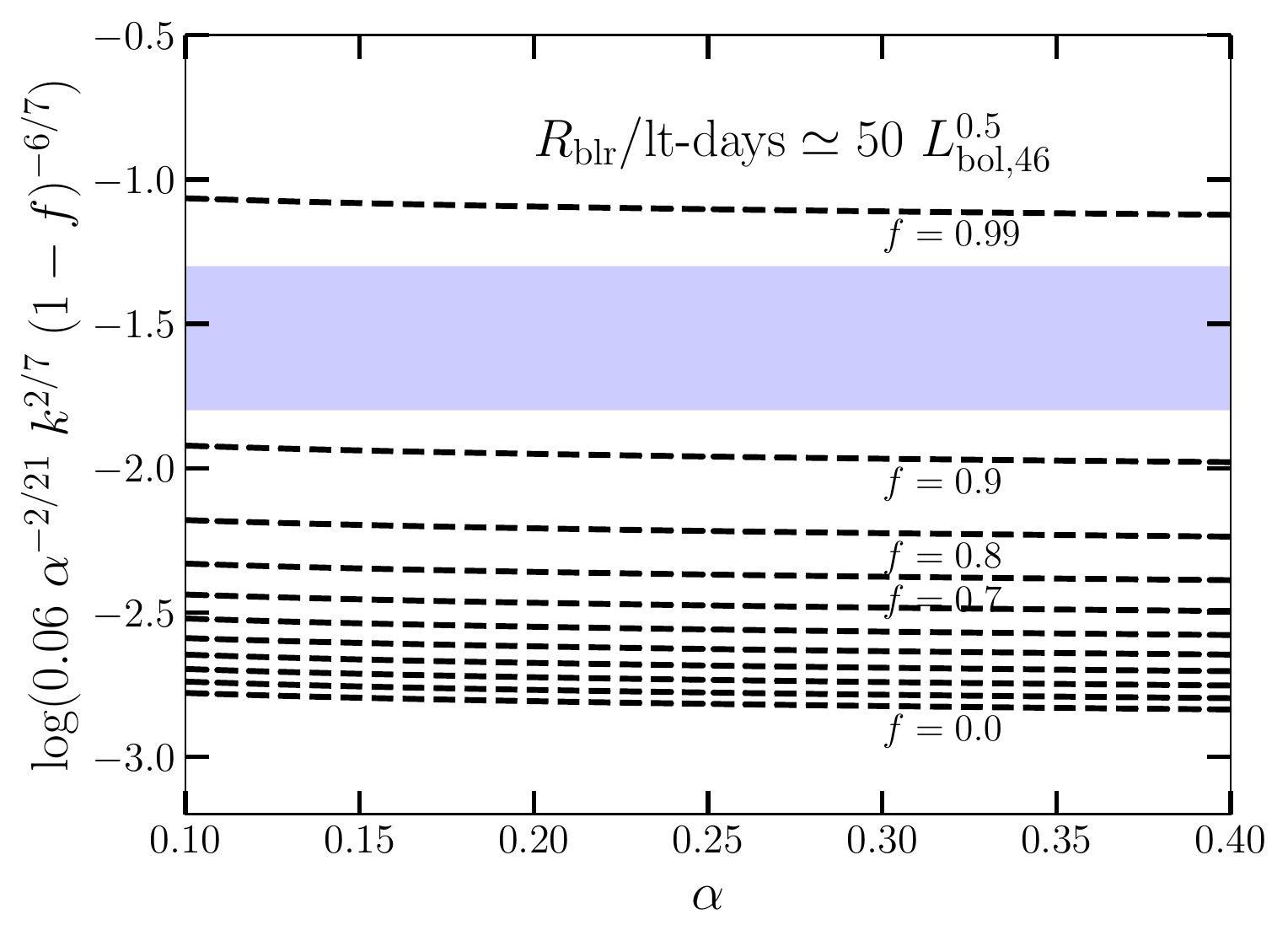}
\caption{The logarithm of the normalization of equation~(\ref{lxlow}) as a function of the accretion disc viscosity, $\alpha$. The factor $\kappa$ is fixed to $1.3\times10^{-6}$, which corresponds to a broad line region size of about $1.3\times10^{17}$ cm ($\sim50$ light days) and $L_{\rm bol}=10^{46}$ erg s$^{-1}$. The dashed black lines represent different values of $f$ in the range from 0 to 0.99. The blue shaded area indicates the observed range of the normalization of the $\Lx-\Lo-\w$ plane in a sample of $\sim500$ quasars. Figure based on data derived from Lusso \& Risaliti (2017).}\label{fig:normlolx}
\end{figure}

Figure~\ref{fig:normlolx} shows how the normalization of equation~(\ref{lxlow}) changes as a function of $\alpha$, $\kappa$, and $f$. We fixed the $\kappa$ factor to $1.3\times10^{-6}$, which corresponds to a broad line region size of about $1.3\times10^{17}$ cm ($\sim50$ light days) and a bolometric luminosity of $10^{46}$ erg s$^{-1}$, typical of quasars (e.g. \citealt{2005ApJ...629...61K}). The viscosity parameter varies from 0.1 to 0.4 \citep{2007MNRAS.376.1740K}, whilst $f$ ranges in the interval 0--0.99. Although the normalization of the relation dsiplays a large scatter, which can be a factor of $\sim2$ in logarithm (i.e. orders of magnitude), the data suggest a value of $f$ in the range 0.9--0.95.

In the future, by studying larger sample of quasars with $M_{\rm BH}$ and accretion rate values (which are all related to $f$), we will compare our results on $f$ with the expectations from numerical simulations (e.g. \citealt{2014ApJ...796..106J}) and other theoretical models (e.g. \citealt{2002MNRAS.332..165M,2009MNRAS.394..207C,2016ApJ...833...35L}), especially those concerning the physical state and extent of the broad line region (e.g. Failed Radiatively Accelerated Dusty Outflows, FRADO, \citealt{2017ApJ...846..154C}). 

\section{Conclusions}
Our modified $\Lx-\Lo$ relationship in quasars, which takes into account the full-width half-maximum of the quasar emission line (i.e. $\Lx\propto\Lo^{\hat\gamma}\w^{\hat\beta}$), has an observed dispersion of $\sim$0.2~dex over $\sim$3~orders of magnitude in luminosity and indicates that there is a good ``coupling" between the disc, emitting the primary radiation, and the hot-electron {\it corona}, emitting X--rays. 

We interpreted such relation through a simple (but physically motivated) model based on the ones presented by \citet{1994ApJ...436..599S} and \citet{2002MNRAS.332..165M}, where a geometrically thin, optically thick accretion disc is coupled with a uniform hot plasma. 
We assumed that the bulk of the corona emission is mainly powered by the accretion disc and it is located at the transition radius ($r_{\rm tr}$) where the gas pressure equates the radiation pressure in the disc. 
Assuming a broad line region size function of the bolometric luminosity as $R_{\rm blr}\propto L_{\rm bol}^{0.5}$ we have that $M_{\rm BH} \propto \dot{M}/\dot{M}_{\rm Edd} \w^4$, which leads to the final relation $\Lx\propto\Lo^{4/7} \w^{4/7}$. Such a relation is remarkably consistent with the fit obtained from a sample of 550 optically selected quasars from SDSS DR7 cross matched with the XMM--{\it Newton} catalogue 3XMM-DR6. The toy model we presented, although simplistic, is capable of making robust predictions on the X--ray luminosities (at a given \rev{ultraviolet} emission and $\w$) of unobscured/blue quasars, and puts observational constraints on the fraction of accretion power released in the hot plasma in the vicinity of the accretion disc, $f$, as a function the broad line region size, from the observed normalization of $\Lx-\Lo-\w$ plane. The latter result will provide a new vantage point for the next generation of semi-analytical and magneto-hydrodinamical simulations investigating on the physical link between the accretion disc and the X-ray corona.

The proposed relation $\Lx\propto\Lo^{4/7} \w^{4/7}$ does not show significant evolution with time in the redshift range covered by our data, $z=0.3-2.2$, and thus it can be employed as a cosmological indicator to estimate cosmological parameters (e.g. $\Omega_{\rm M}$, $\Omega_{\Lambda}$).



\section*{Funding}
EL is supported by a European Union COFUND/Durham Junior Research Fellowship (under EU grant agreement no. 609412).
This work has been supported by the grants PRIN-INAF 2012 and ASI INAF NuSTAR  I/037/12/0.

\section*{Acknowledgments}
For all catalogue correlations we have used the Virtual Observatory software TOPCAT \citep{2005ASPC..347...29T} available online (http://www.star.bris.ac.uk/$\sim$mbt/topcat/).
This research has made use of data obtained from the 3XMM XMM--Newton serendipitous source catalogue compiled by the ten institutes of the XMM--Newton Survey Science Centre selected by ESA. This research made use of matplotlib, a Python library for publication quality graphics \citep{2007CSE.....9...90H}.


\bibliographystyle{frontiersinHLTH&FPHY} 
\bibliography{bibl}






\end{document}